\documentclass[a4paper,10pt]{article}
\usepackage[utf8]{inputenc}

\usepackage{publication}
\usepackage[backend=biber,sorting=none]{biblatex}
\usepackage{amsmath} 
\usepackage{graphicx} 
\title{3D space-variant modal deconvolution with computed point spread functions}

\author[1]{Jakub Czuchnowski}
\contact{Correspodning author: Jakub Czuchnowski - jczuchno@bu.edu}
\affil[1]{Department of Biomedical Engineering, Boston University, Boston, MA, 02215, USA}
\affil[3]{Photonics Center, Boston University, Boston, MA, 02215, USA}
\affil[2]{Department of Biology and The Picower Institute for Learning and Memory, MIT, Cambridge, MA, 02139, USA}
\author[1]{Chuan Li}
\author[2]{Hongli Ni}
\author[2]{Brandon Weissbourd}
\author[1,3]{Jerome Mertz}

\date{\today}

\begin{document}

\maketitle

\begin{abstract}
    Deconvolution is the most widely used aberration correction technique in  microscopy, however most techniques assume that the aberrations are the same for each point in the image, which is rarely true. Methods for tracking spatially varying aberrations require burdensome calibration or computation, or require symmetries in the aberration patterns. Here, we expand on existing modal deconvolution methods to demonstrate 3D fluorescence deconvolution in imaging systems that exhibit no simple symmetry. Our method is based on a space-variant generalization of Richardson-Lucy deconvolution that makes use of ZEMAX\textsuperscript{\textregistered}-derived point spread functions without the requirement of guide stars or calibration measurements. We validate the performance of our method by applying it to snapshot multiplane imaging of both bead samples and biological specimens, and show that modal decomposition is a practical solution for deconvolving spatially varying aberrations that do not display clear symmetries.
\end{abstract}



\begin{multicols}{2}

\section{Introduction}
In many cases, the practical performance limits of imaging systems are set by optical aberrations, which degrade image resolution and signal-to-noise ratio (SNR). For this reason, much effort in optical engineering is put into aberration correction, whether by advanced lens design \cite{kingslake2009lens} or by introduction of active elements \cite{booth2007adaptive,hampson2021adaptive} that can correct aberrations during imaging. However, ultimately due to cost and added complexity, deconvolution remains the most common aberration correction method. While several deconvolution algorithms exist, the problem of addressing spatially varying aberrations (where the point spread function [PSF] varies depending on its location within the object) still poses significant challenges. Because most deconvolution algorithms assume space invariance, approaches to address spatially variant aberrations have involved splitting images into sub-regions that are assumed to be space invariant either blindly \cite{trussell1978image} or adaptively \cite{lin1987piecewise,lohmann1965space}, or interpolating PSFs \cite{denis2011fast,nagy1998restoring}. These approaches require considerable efforts to calibrate and characterize hundreds to thousands of PSFs, and add significant computational overhead, making them experimentally impractical. Recent work has demonstrated that spatially variant aberrations can be addressed efficiently using a method of ring deconvolution \cite{kohli2025ring} if their underlying aberration patterns exhibit rotational symmetry. In many cases, however, this symmetry can be broken, for example by the inclusion of non-rotationally symmetric optical elements such as microlens arrays \cite{levoy2006light} or image splitting prisms \cite{xiao2020high}. 

To address such cases, we expand on a previously explored approach of decomposing spatially varying aberrations into spatially invariant modes and incorporating them into a Richardson-Lucy (RL) deconvolution algorithm  \cite{lauer2002deconvolution,turcotte2020deconvolution}. While these methods do not approach the computational speed of ring deconvolution, they are still significantly faster than image subdivision or PSF interpolation, especially for non-compact PSFs (such as a widefield PSF) because the computational complexity does not increase linearly with the number of characterized PSFs but only with the number of significant modes required to approximate the PSFs \cite{turcotte2020deconvolution}. Most importantly, these methods are capable of addressing aberration patterns lacking rotational symmetry even in datasets displaying uneven spatial sampling. However, these methods are not without drawbacks, most notably: 1) the need for thorough characterization of the PSF variations across the imaging volume, which can be both experimentally challenging and time consuming, and 2) RL deconvolution requires positive-definite PSFs, whereas an approximate modal basis with a reduced number of modes does not in general guarantee positivity of the reconstructed PSFs, potentially leading to reconstruction artifacts \cite{lauer2002deconvolution}. In this work, we address these drawbacks and demonstrate, both theoretically and experimentally, that modal decomposition is a practical solution to deconvolving spatially varying aberrations that do not display clear symmetry.

\section{Principle of spatially-variant deconvolution using a modal basis}

Richardson-Lucy deconvolution \cite{lucy1974iterative} is an algorithm that approximates the object distribution by iteratively updating an initial guess ($o_0$):

\begin{equation}
    o_{n+1}=C_no_n
    \label{eq:RL1}
\end{equation}
where $o_n$ is the approximated object distribution at the $n$-th iteration and $C_n$ is the update field:
\begin{equation}
    C_n=\frac{I}{I_n}\otimes \text{PSF}^T
    \label{eq:RL2}
\end{equation}
where $I$ is the recorded image, $\otimes$ denotes a convolution, PSF$^T$ is the flipped PSF [$\text{PSF}^T(u,v,w)=\text{PSF}(-u,-v,-w)$]  and $I_n$ is the image  approximated at the $n$-th iteration:

\begin{equation}
    I_n=o_n \otimes \text{PSF}
    \label{eq:RL3}
\end{equation}

As can be seen, the RL algorithm is based on two convolutions and assumes spatial invariance of the PSF, i.e. that the PSF is not a function of the spatial coordinates of the object $(x,y,z)$ but of the difference between the object and image coordinates:
\begin{equation}
    I(u,v,w)=o(x,y,z)\otimes \text{PSF}(u-x,v-y,w-z)
\end{equation}

\begin{figure*}[ht]
\centering\includegraphics[width=14cm]{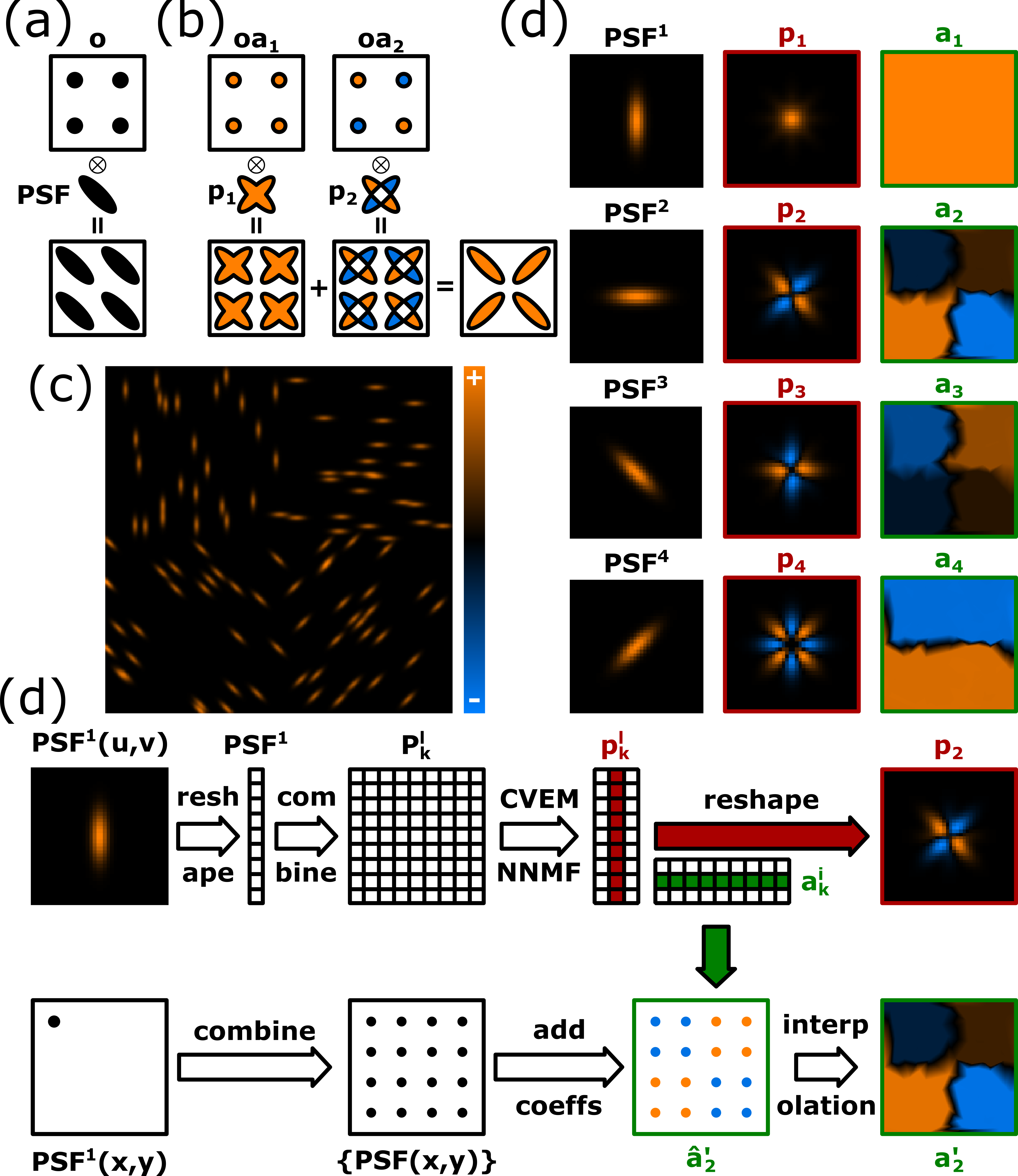}
\caption{(a-b) Conceptual comparison of regular convolution (a) with modal convolution (b). Modal convolution enables efficient convolution of objects with spatially varying PSFs. (c) Simulated image with spatially varying PSFs. (d) Illustration of PSFs, modes (red frame) and coefficient fields (green frame) corresponding to image from (c). (e) Workflow explaining the steps for calculating modes and coefficient fields from a series of PSFs.}
\label{fig:1}
\end{figure*}

In the event the invariance assumption is not met (e.g. because of spatially varying aberrations), the PSF becomes a function of 6 variables, significantly complicating matters:
\small
\begin{equation}
    I(u,v,w)=o(x,y,z)\otimes \text{PSF}(u-x,v-y,w-z,x,y,z)
    \label{eq:I}
\end{equation}
\normalsize
Several potential solutions to this complication have been proposed \cite{lauer2002deconvolution,turcotte2020deconvolution,denis2015fast,flicker2005anisoplanatic,deng2014additive,miraut2012efficient,popkin2010accurate,sroubek2016decomposition,lee2017generalized}. Here, we explore the use of modal convolution [Fig. \ref{fig:1}(a)] by treating the PSF as a separable function and decomposing it into a linear combination of modes ($p_i(u-x,v-y,w-z)$) that are spatially invariant and their corresponding coefficient fields ($a_i(x,y,z)$) that encode the spatial variation of the PSFs [Fig. \ref{fig:1}(b-c)]:
\begin{multline}
    \text{PSF}(u-x,v-y,w-z,x,y,z)=\\
    \sum_i p_i(u-x,v-y,w-z)a_i(x,y,z)
    \label{eq:piai}
\end{multline}
Combining Equations \ref{eq:I} and \ref{eq:piai}, we get: 
\small
\begin{equation}
    I(u,v,w)=\sum_i o(x,y,z)a_i(x,y,z)\otimes p_i(u-x,v-y,w-z)
\end{equation}
\normalsize
which simplifies a spatially variant convolution into a series of spatially invariant ones. Within this framework, Eqs. \ref{eq:RL2} and \ref{eq:RL3} become \cite{lauer2002deconvolution}:

\begin{equation}
    C_n=\sum_i \frac{I}{I_n}a^\tau_i\otimes (p_i)^T
\end{equation}

\begin{equation}
    I_n=\sum_i o_n a_i \otimes p_i
\end{equation}
which together with Eq. \ref{eq:RL1} form the basis of our variant Richardson Lucy (vRL) deconvolution method. We note here that $a_i \not\equiv a_i^\tau$, which is a subtle difference that will be explained below.

\section{Determining the modes and coefficient fields}

As mentioned above, fully characterizing the spatial variations of the PSF would require a very dense sampling of the imaged volume, which is generally prohibitive due to technical reasons \cite{lauer2002deconvolution,kohli2025ring}. A potential solution to this is to sparsely sample the volume and use interpolation to achieve a continuous coefficient field ($a_i$, Fig. \ref{fig:1}(d)). Such sparse sampling generates a dataset of discretized PSFs together with their spatial localization:
\small
\begin{equation}
\{\text{PSF}_k(u-x_k,v-y_k,w-z_k,x_k,y_k,z_k)\}=\{\text{PSF}_k^{u,v,w}\}. 
\end{equation}
\normalsize
where the superscripts $u,v,w$ are the pixel indices in the discretized PSF. By reshaping the PSF matrix into linear vector form ($\text{PSF}_k^{u,v,w}=\text{PSF}_k^l$, where $l$ is the linear index), we can express the full dataset as a matrix: 
\begin{equation}
    P_k^l\overset{def}{=}\text{PSF}_k^l
\end{equation}

Our goal here is to achieve a modal approximation of the measured PSFs that can be represented by:

\begin{equation}
    P^l_k \approx \sum_i p_i^l\, \hat{a}^i_k
\end{equation}
such that the residual between the modeled and measured PSFs is minimized. Several approaches are available for this, out of which we describe two. In a first approach, we can use the eigenmodes ($e_i^k$) of the covariance matrix  ($\text{CV}^{k1}_{k2}$) (CVEM) to calculate the modes ($p_i$) and coefficient fields ($a_i$) \cite{lauer2002deconvolution}:

\begin{equation}    \text{CV}^{k1}_{k2}=\text{cov}(\text{PSF}_{k1}^l,\text{PSF}_{k2}^l)
\end{equation}

\begin{equation}
    e_i^k=\text{eig}(\text{CV})
\end{equation}
where $eig()$ denotes extracting the eigenvectors, which in turn are used to calculate the spatial modes:
\begin{equation}
    p_i^l=\sum_k e_i^k \, \text{PSF}_k^l
\end{equation}
and discretized coefficient fields:
\begin{equation}
    \hat{a}^i_k=(p_i \cdot PSF_k)
\end{equation}
where $(\ \cdot\ )$ denotes a vector dot product.

The advantage of this method lies in its ability to efficiently encode the spatial variability of the PSF into a small number of modes. The disadvantage lies in the fact that, since the generated modes have both positive and negative values, if an insufficient number of modes is selected for deconvolution the effective PSF models might contain negativities that prevent the algorithm from converging \cite{lauer2002deconvolution}.  In such situations, an alternative approach using non-negative matrix factorization \cite{wang2012nonnegative} (NNMF) can achieve a similar (albeit inferior, as shown below) deconvolution results without introducing negativities into the PSF models:

\begin{equation}
    \text{NNMF}(P^l_k) = [p_i^l,\hat{a}^i_k]
\end{equation}

The discreet coefficient fields ($\hat{a}^i_k$) can then be used to calculate continuous approximations of the actual coefficient fields via interpolation:
\begin{equation}
    a'_i(x,y,z)=\text{interp}(\hat{a}^i_k,x_k,y_k,z_k,z,y,z)
\end{equation}

A final step involves normalizing the coefficient field to ensure that the reconstructed PSFs are normalized:
\begin{equation}
    a_i(x,y,z)=\frac{a'_i}{\sum_ja'_j\otimes  p_j}
\end{equation}
We note that $a_i \not\equiv a_i^\tau$ because the normalization depends on the forms of $p_i$ used for the deconvolution:
\begin{equation}
    a_i^\tau(x,y,z)=\frac{a'_i}{\sum_ja'_j\otimes  (p_j)^T}
\end{equation}

\begin{figure*}[ht]
\centering\includegraphics[width=14cm]{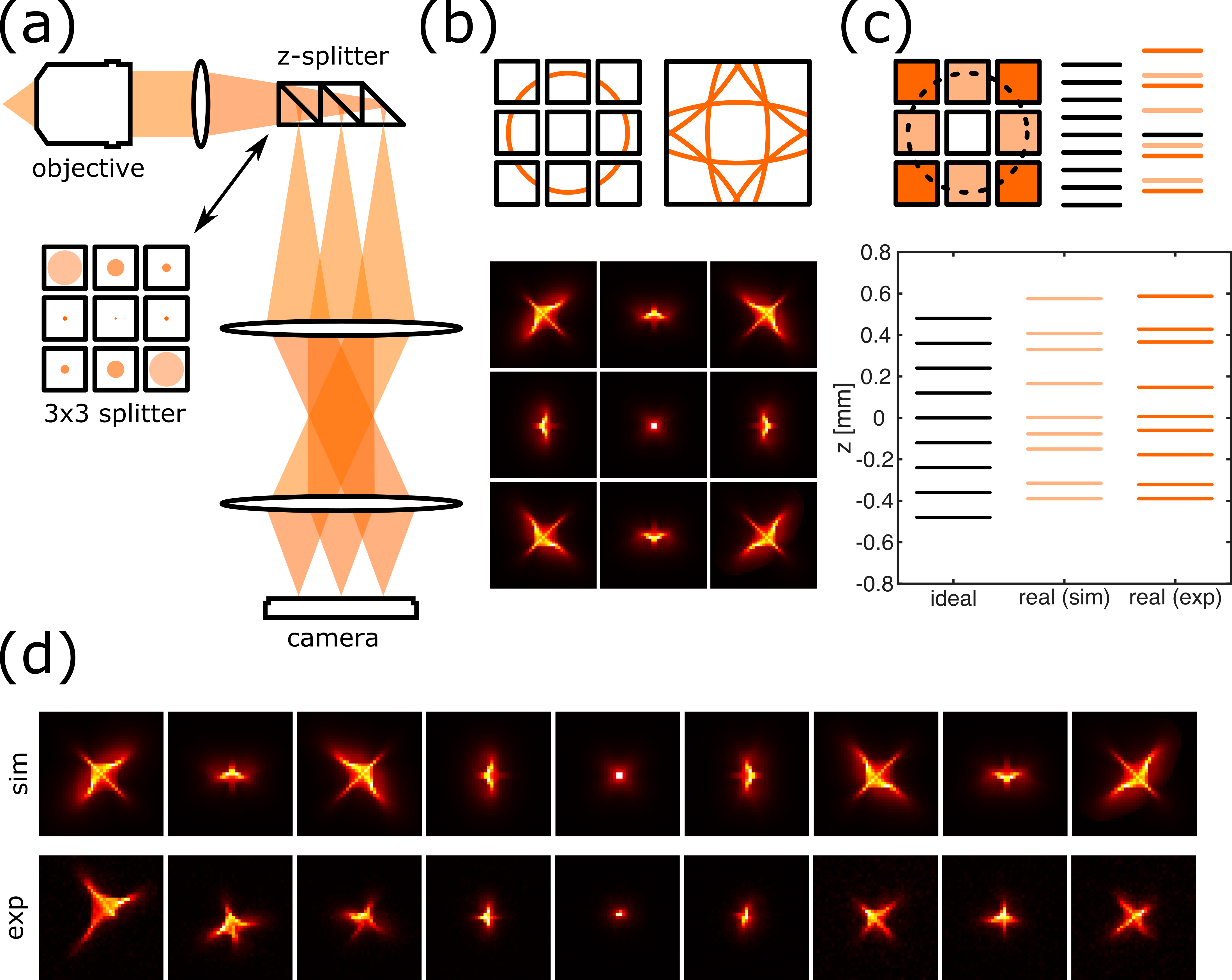}
\caption{(a) Schematic of the multi-z widefield microscope used in this work. (b) Effect of the z-splitter prism on breaking the rotational symmetry of the aberrations in the reconstructed 3D volume.  (c) Effect of field curvature on the spacing of the 9 z-planes compared to the design spacing. (d) Comparison between maximum intensity projections of experimental and simulated PSFs.}
\label{fig:2}
\end{figure*}

\section{Demonstration system}


For our demonstrations, we choose to use a widefield multiplane microscope with a z-splitting prism \cite{xiao2020high} [Fig. \ref{fig:2}(a)]. This microscope system is essentially composed of two parts: the pre-splitter part, which is a standard widefield microscope, and a post-splitter part that relays the multiple image planes to a single camera.  The pre-splitter part can suffer from rotationally symmetric aberrations, which can be corrected by conventional approaches. However, the introduction of the image splitting prism breaks the rotational symmetry by displacing the image planes into different off-axis positions. As a result, while the splitter itself does not introduce significant aberrations, the post-splitter relay can introduce noticeable astigmatism [Fig. \ref{fig:2}(b)]. This is particularly true if no efforts are made to control these aberrations by proper optical design (e.g. by the use of Pl\"ossl or telecentric lenses), as was intentionally the case here. Because of the system geometry, the astigmatism additionally rotates around a differently located axis for each image plane [Fig. \ref{fig:2}(b)]. This generates a 3D aberration pattern with no simple symmetry where aberrations experienced by the PSF not only vary in all three dimensions, but also different z-planes of the same PSF effectively experience different aberrations. Moreover, due to the field curvature introduced by the post-splitter relay, the spacing between different focal planes can deviate strongly from the designed uniform spacing [Fig. \ref{fig:2}(c)]. For all these reasons, a widefield multiplane microscope (improperly designed) thus constitutes a perfect testbed for the demonstration of our 3D-vRL framework.

The experimental data presented in this paper were collected using two different setups following the same overall design principles. The first was a large-volume system equipped with a $10\times$ objective (Olympus, UPlanFL N, $0.3$ NA) with an effective magnification of $2.8\times$, leading to an effective pixel size of $2.34 \, \mu$m in the sample plane and a $936 \times 936 \times 960 \, \mu$m field of view (PCO Edge 4.2 LT, $2048 \times 2048$ pixels, $6.5 \, \mu$m pixel size). The second was a high-speed system featuring a $10\times$ objective (Zeiss, Objective Plan-Apochromat $10\times/0.45$) with an effective magnification of $10.4\times$, leading to an effective pixel size of 625 nm in the sample plane and a $312 \times 312 \times 50 \,\mu$m field of view (Teledyne Kinetix, $2200 \times 2200$  pixels, $6.5 \,\mu$m pixel size).

\section{Computational generation of PSFs}

A major challenge in implementing spatially varying deconvolution methods lies in the need for a thorough characterization of the spatial variation of the PSF. In practice, this can be done experimentally by scanning a sample of sub-resolution beads across the field of view and fitting PSF models to the measured images. While laborious, this is readily feasible in a 2D case since only minimal scanning is required when using a dense sample of beads. However, in a 3D case with compact PSFs (e.g. confocal) this approach becomes more onerous as the sample must be scanned in 3D; and in the case of non-compact PSFs (e.g. widefield) it becomes almost unfeasible since only a single bead would need to be scanned throughout the entire volumetric field of view to avoid out-of-focus overlap of PSFs. Alternatively, one could attempt to derive the aberrated PSFs theoretically. While this could provide a calibration with arbitrarily dense PSF sampling, it becomes progressively more difficult as the system complexity grows, to the point of also becoming impractical in many experimental applications. In this work, we opted for a hybrid approach that combines Monte-Carlo simulated measurements and system modeling within the framework of an optical design software. 

Because our system suffers primarily from geometric aberrations, we used a ray tracing approach using ZEMAX\textsuperscript{\textregistered} to approximate the aberrated PSFs with good accuracy [Fig. \ref{fig:2}(d)]. The advantage of using ZEMAX\textsuperscript{\textregistered} lies in its vast library of models for most off-the-shelf optical components from all major vendors, facilitating a broad applicability of this method of PSF modeling. For this particular work, we simulated an evenly spaced grid of $11 \times 11$ PSFs for each of the 9 planes of the z-splitter prism, yielding a total of $1089$ different PSFs. Because of the significant field curvature present in the system, we chose to deconvolve based on the PSFs simulated in the real planes of the microscope as opposed to attempting to deconvolve back to the ideal planes. This allowed us to fully exploit the in-focus light information present in the datasets and achieve optimal image reconstruction quality. 

\begin{figure*}[ht]
\centering\includegraphics[width=14cm]{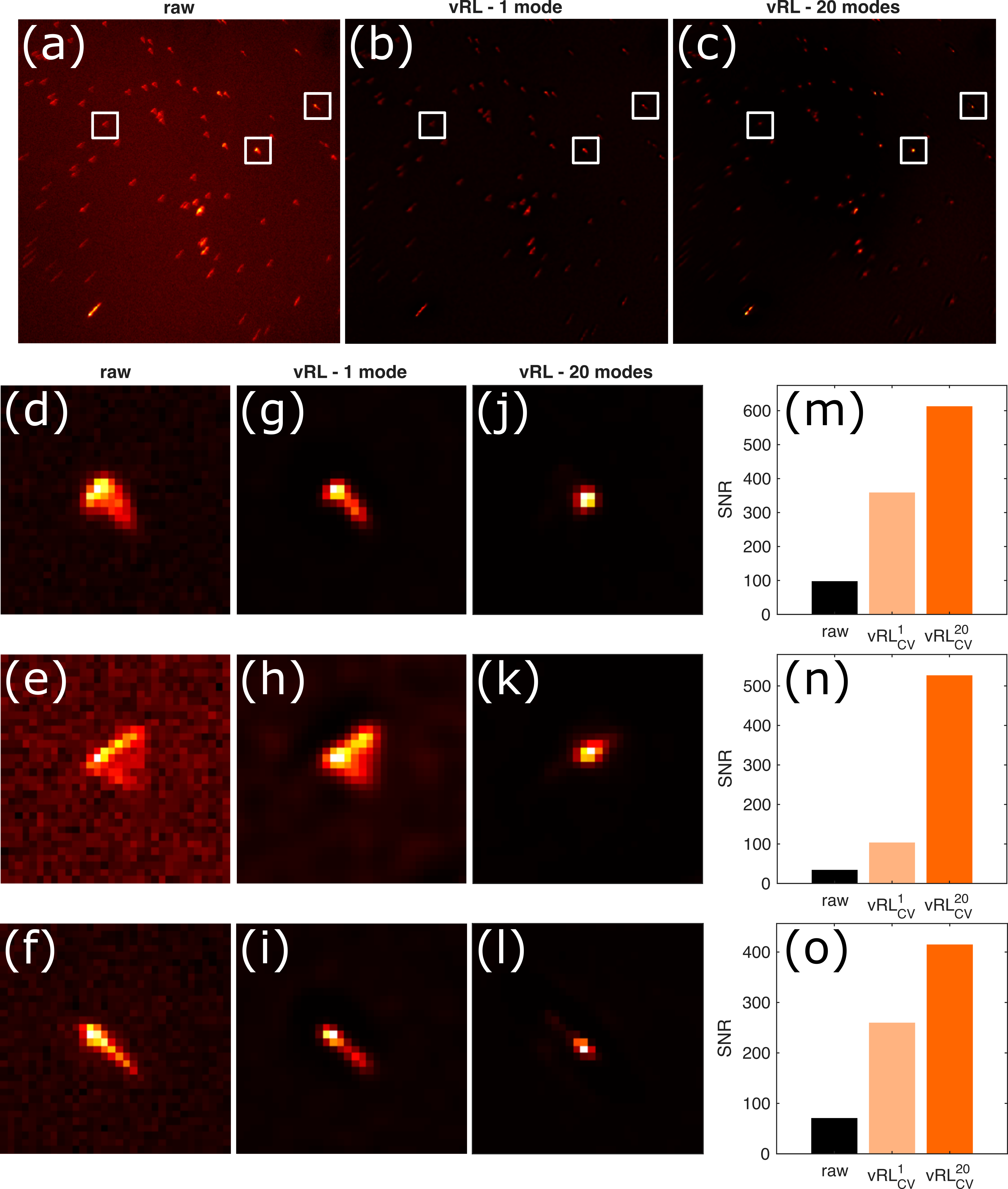}
\caption{(a-c) Images of bead samples deconvolved using a 1-mode (b) and 20-mode vRL (c) for 100 iterations with CV modes (the FOV is $0.936 \times 0.936$ mm). (d-l) Zoom-ins of individual PSFs, comparing raw (d-f), deconvolved with 1 mode (g-i) and 20 modes (j-l). (m-o) Estimated SNR for PSF in panels (d-l).}
\label{fig:3}
\end{figure*}

\section{Experimental results}

We first validate the performance of our approach using 1 $\mu m$ fluorescent beads on a slide [Fig. \ref{fig:3}(a,d-f)]. We note that while the sample itself is 2D, we collect a full 3D dataset and perform 3D deconvolution, which allows us to improve image SNR by exploiting the recorded out-of-focus light. We compare deconvolving using only the first CV mode (corresponding to the average PSF) and using a reasonably large number of modes (in this case $20$). As can be observed, even using a single CV mode does yield some improvement to the images [Fig. \ref{fig:3}(b)] and leads to increased SNR (here approximated by the intensity mean of the 4 brightest pixels divided by the intensity standard deviation at the edge of the PSF ROI) [Fig. \ref{fig:3}(m-o)]. This improvement stems from several effects, such as the refocusing of out-of-focus signals from other imaging planes back to the focal plane, the suppression of background noise, and the enhancement of PSF compactness. However, the resulting PSFs are still remain strongly aberrated, and their overall distorted shape does not change significantly [Fig. \ref{fig:3}(g-i)]. On the other hand, deconvolution with $20$ modes yields not only a further improvement in SNR [Fig. \ref{fig:3}(m-o)] but also a significant reduction in the spatially dependent distortion of the PSF shape, recovering in many cases a highly symmetric distribution [Fig. \ref{fig:3}(j-l)].

\begin{figure*}[ht]
\centering\includegraphics[width=14cm]{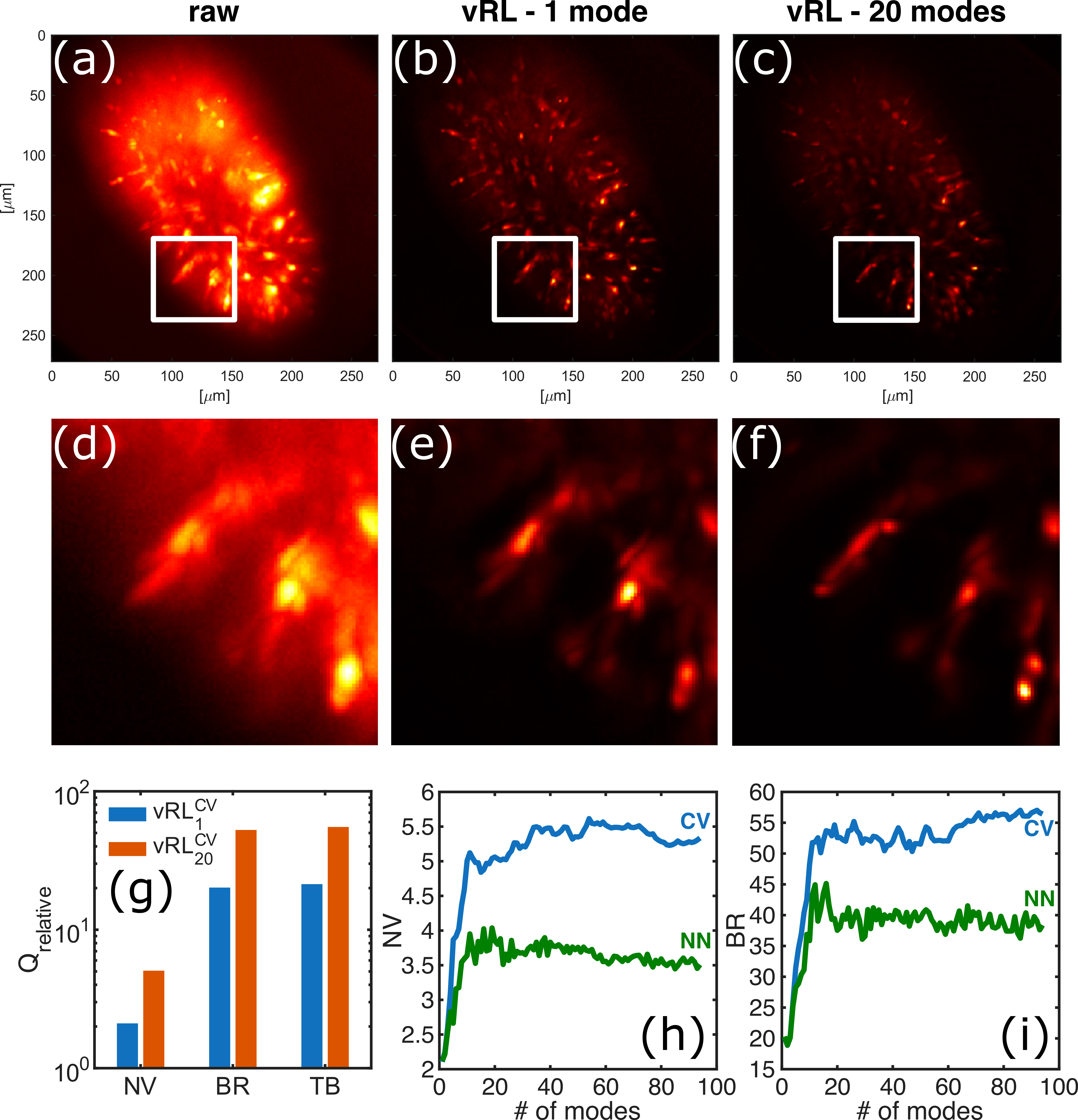}
\caption{(a-f) Maximum intensity z-projections of 3D volumes of a fixed larva sample. From the left: the raw widefield image (a), a vRL deconvolution using a single CV mode (b), a vRL deconvolution using 20 CV modes (c) and their corresponding zoom-ins (d-f). (g) Quantification of relative image quality for datasets from (a) compared to the raw data using Normalized Variance, and Brenner-Tennenbaum methods. (h-i) Comparison of Normalized Variance (h) and Brenner-Tennenbaum (i) image quality metrics for the datasets from (a-c) as a function of number of modes used for vRL with the CVEM (blue) and NNMF (green) methods.}
\label{fig:4}
\end{figure*}

To further test our approach on more complex samples we proceed to deconvolve images of the larval stage from \textit{Clytia hemisphaerica}, a species of jellyfish labeled via transgenic expression of mCherry\cite{weissbourd2021genetically}  [Fig. \ref{fig:4}(a,d)]. Here, as well, we compare $20$-mode vRL [Fig. \ref{fig:4}(c,f)] with single-mode vRL, which we treat as a benchmark [Fig. \ref{fig:4}(b,e)]. We estimate image quality with widely used sharpness metrics \cite{treeby2011automatic,czuchnowski2022comparing} including Normalized Variance as well as Brenner and Tennenbaum gradient methods. We observe that even single-mode vRL yields improvements in relative quality [Fig. \ref{fig:4}(g)] as compared to the raw data (in agreement with the increased SNR obtained with sub-diffraction beads, [Fig. \ref{fig:3}(m-o)]) mainly due to deblurring and removal of unfocused light. However, increasing the number of modes to $20$ further improves all metrics significantly,  confirming that our method of modal deconvolution is capable of correcting spatially variant aberrations even for complex biological samples.

We then proceed to analyze the dependence of the resulting image quality on the number of modes used for deconvolution and observe that the images experience a sharp improvement in estimated quality when increasing the number of modes from 1 to $\sim$10 [Fig. \ref{fig:4}(h,i)]. However, further increasing the number of modes yields diminishing returns which eventually plateau, suggesting that about $10$ modes exhaust most of the spatial variability of the PSFs in our system. 

\section{Conclusion}

Addressing spatially varying aberrations still presents a challenge, especially in optical systems without a clear symmetry in their aberration patterns. Our work demonstrates that modal decomposition is capable of addressing this problem in a computationally efficient manner. We show that over $1000$ individual PSFs can be represented by just $\sim10$ modes while still yielding high quality deconvolution. Our modal decomposition method does slow the overall deconvolution computation time in proportion to the number of modes used for deconvolution. However, because all the convolutions in a single iteration are independent, in principle they can be parallelized to reduce the computational time depending on the computer specifications and the size of the deconvolved images (assuming sufficient GPU memory). We emphasize that while we chose to demonstrate our method with an improperly designed multi-z widefield microscope, its range of applicability is much broader than this, allowing aberration correction in both 2D and 3D imaging configurations with arbitrarily spaced planes. Normally, one of the limitations in addressing spatially varying aberrations is the need for a detailed characterization of the PSFs of the system, which is especially challenging when using widefield microscopes with non-compact PSFs (particularly in the defocused planes). Here, however, we demonstrate the possibility of effective 3D widefield deconvolution without the measurement of a single PSF, by making use of easily accessible optical design software such as ZEMAX\textsuperscript{\textregistered}. Such a possibility of measurement-free space-variant deconvolution is, to our knowledge, unprecedented and promises to be of general utility to the imaging community.   

\section*{Funding}
This work was funded by the National Science Foundation (EEC-1647837) and the National Institutes of Health (R01NS116139, R01GM160992).

\section*{Disclosures}
The authors declare no conflicts of interest.

\section*{Data availability} Data underlying the results presented in this paper are not publicly available at this time but may be obtained from the authors upon reasonable request.

\end{multicols}

\printbibliography
\end{document}